\documentclass[12pt]{article}

\usepackage{arxiv}
\usepackage[utf8]{inputenc} 
\usepackage[T1]{fontenc}    
\usepackage{hyperref}       
\usepackage{url}            
\usepackage{booktabs}       
\usepackage{amsfonts}       

\usepackage{graphicx}
\usepackage{doi}
\usepackage[dvipsnames]{xcolor}
\usepackage{amsmath, amssymb}
\usepackage{cleveref}    

\usepackage{subcaption}
\usepackage{bm}

\usepackage[version=4]{mhchem}
\usepackage{cleveref}
\crefrangelabelformat{equation}{#3#1#4--#5#2#6}
\crefname{equation}{Eq.}{Eqs.}   
\usepackage{authblk}
\usepackage[backend=biber,style=nature, url=false, isbn=false]{biblatex}
\AtEveryBibitem{\clearfield{month}}
\AtEveryBibitem{\clearlist{language}}

\usepackage[mathlines]{lineno}
\usepackage{enumitem}
\usepackage{comment}
\usepackage{multicol}
\usepackage{setspace}

\usepackage{siunitx}
\DeclareUnicodeCharacter{03BC}{\textmu}
\DeclareUnicodeCharacter{2212}{-}
\DeclareSIUnit\bar{bar}
\DeclareSIUnit{\pixel}{px}
\title{The Refractive Index of Gallium Antimonide}

\newbox{\orcid}\sbox{\orcid}{\includegraphics[scale=0.06]{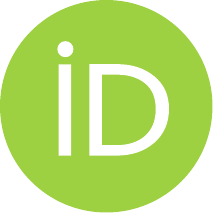}} 

\author[1,2]{%
	\href{https://orcid.org/0009-0002-3052-3083}{%
    \usebox{\orcid}\hspace{1mm}Ulrich~Galander%
    \thanks{\texttt{ulrich.galander@univie.ac.at}}%
    }%
}

\author[3,4]{%
	\href{https://orcid.org/0000-0003-3741-7160}{%
    {\usebox{\orcid}\hspace{1mm}Nicolas~Huwyler}}%
}

\author[1]{%
	\href{https://orcid.org/0000-0001-5169-7328}{%
    {\usebox{\orcid}\hspace{1mm}Mirela Encheva}}%
}

\author[3,5]{%
	\href{https://orcid.org/0000-0002-1432-8539}{%
    {\usebox{\orcid}\hspace{1mm}Matthias~Golling}}%
}

\author[1]{%
	\href{https://orcid.org/0000-0003-2879-1564}{%
    \usebox{\orcid}\hspace{1mm}Oliver H.~Heckl%
    \thanks{\texttt{oliver.heckl@univie.ac.at}}%
    }%
}

\affil[1]{Optical Metrology, Faculty of Physics, University of Vienna, Boltzmanngasse 5, 1090 Vienna, Austria}
\affil[2]{Vienna Doctoral School in Physics, University of Vienna, Boltzmanngasse 5, 1090 Vienna, Austria}
\affil[3]{ETH Zurich, Department of Physics, Institute for Quantum Electronics, Auguste-Piccard-Hof 1, 8093 Zurich, Switzerland}
\affil[4]{Université de Neuchâtel, Institut de Physique, Laboratoire Temps-Fréquence, Avenue de Bellevaux 51, 2000 Neuchâtel, Switzerland}
\affil[5]{RhySearch, Optical Coating \& Characterization Lab, Werdenbergstrasse 4, 9471 Buchs, Switzerland}

\renewcommand{\headeright}{Research Article}
\renewcommand{\undertitle}{Research Article}

\begin{document}
\maketitle
\setcounter{footnote}{0}
\begin{abstract}

Gallium antimonide (GaSb) is a key material for near- and mid-infrared photonics, enabling high-performance laser architectures and detectors. Design and simulation of such devices depend on accurate optical material data, especially the complex refractive index $n^*_{\text{GaSb}} = n_{\text{GaSb}} +ik_{\text{GaSb}}$, consisting of the real part $n_{\text{GaSb}}$ (refractive index) and the imaginary part $k_{\text{GaSb}}$ (extinction coefficient). However, GaSb refractive index values are based either on theoretical models, typically informed by legacy experimental data, or on experimental measurements without quantified uncertainties. This limits their reliability for state-of-the-art devices. Here, we present measurement results of $n^*_{\text{GaSb}}$ in the near- to mid-infrared range from \SIrange{1}{3.1}{\micro \metre} with a relative uncertainty <\num{7.8e-5} for $n_{\text{GaSb}}$, and <\num{2.0e-3} for $k_{\text{GaSb}}$. As a side result of our method, we also report $n_{\text{AlAsSb}}$ for aluminium arsenide antimonide ($\mathrm{AlAs_{0.08}Sb_{0.92}}$) with a relative uncertainty <\num{3.9e-4}. Our results are based on two complementary measurements on a GaSb/AlAsSb-based heteroepitaxial structure under controlled environmental conditions: photometric transmission and layer-thickness analysis by cross-sectional scanning electron microscopy. We simultaneously retrieve the refractive indices of the two materials by fitting a Sellmeier equation and a theoretical dispersion model by Djuri\v{s}i\'c \textit{et al.}~\cite{djurisic_modeling_2000}.
The uncertainties of $n^*_{\text{GaSb}}$ and $n_{\text{AlAsSb}}$ are quantified using a Monte Carlo-based approach. Our results provide accurate complex refractive index values for GaSb, which are vital to advance photonics-related technologies in the near- and mid infrared spectral region.

\end{abstract}
\keywords{refractive index \and semiconductor \and gallium antimonide \and optical metrology}

\section{Introduction}\label{sec:1.Introduction}

Semiconductors fundamentally changed modern technology through their widespread use in applications such as solar cells, diodes, and transistors in integrated circuits, thus making them a crucial component in many electronic devices. Consequently, a thorough understanding of their properties is essential to enable further technological advances in these areas. 

Among semiconductors, GaSb is of particular interest because of its narrow band gap of $\approx$\SI{0.72}{\eV} ($\approx$\SI{1.7}{\micro \metre}), which makes it suitable for a variety of applications. For example, GaSb is used in thermophotovoltaic cells, demonstrating high efficiencies~\cite{bett_gasb_2003}, and Huang~\textit{et al.}~\cite{kwei-wei_huang_active_2013} developed an InAs/GaSb semiconductor-based type-II superlattice for imaging in the near-infrared (NIR) (\SIrange{1.7}{3}{\micro \metre}) and mid-infrared (MIR) (\SIrange{3}{5}{\micro \metre}) spectral regions. Furthermore, GaSb-based heteroepitaxial structures are essential for active optoelectronical devices operating in the molecular fingerprint region. Examples include vertical-cavity surface-emitting lasers (VCSELs)~\cite{cerutti_gasb-based_2009}, superluminescent diodes~\cite{zia_high-power_2019}, quantum cascade lasers (QCLs)~\cite{joullie_gasb-based_2003,yasuda_growth_2024}, and vertical external-cavity surface-emitting lasers (VECSELs)~\cite{gaulke_optically_2025}, as well as passive elements like the semiconductor saturable absorber mirror (SESAM)~\cite{alaydin_bandgap_2022,schuchter_composition-controlled_2025}. 

To advance the development of these applications, precise knowledge of the optical properties of GaSb is vital. Among these optical properties, the complex refractive index $n^*$ is fundamental because it governs light propagation and light-matter interaction. For GaSb, several experimental~\cite{aspnes_dielectric_1983, munoz_uribe_near-band-gap_1996,ferrini_optical_1998,roux_mid-infrared_2015,wasiak_absorption_2018} and theoretical studies~\cite{adachi_optical_1989,paskov_refractive_1997,djurisic_modeling_2000,linnik_calculations_2002} have determined $n^*$.
Typically, theoretical studies derive parameter-based models for the electric permittivity (connected to the complex refractive index via Eq.~\ref{equ:refractive_dielectric}) and use published experimental data to evaluate the parameters, whereas experimental studies primarily use spectroscopic ellipsometry (SE) to obtain the refractive index due to its non-destructive measurement approach and ability to perform \textit{in situ} measurements during thin-film growth. However, these studies have certain shortcomings: 
(i) Most theoretical models rely either on the experimental datasets from Aspnes~\textit{et al.}~\cite{aspnes_dielectric_1983}, which cover photon energies from \SIrange{1.5}{6}{\eV} ($\approx$\SIrange{0.206}{0.826}{\micro \metre}), or on data from Ferrini~\textit{et al.}~\cite{ferrini_optical_1998}, which include the GaSb bandgap region as well, making these two datasets the main experimental sources.
(ii) The experimental studies investigate different sample types but report only limited sample details, making meaningful comparisons difficult. For example, Aspnes~\textit{et al.}~\cite{aspnes_dielectric_1983} measured p-doped (\SI{1.5e17}{\per\centi\meter\cubed}) bulk GaSb oriented along the $\langle 1 1 1 \rangle$ crystal axis; Ferrini \textit{et al.}~\cite{ferrini_optical_1998} measured a molecular beam epitaxy (MBE) grown single layer of GaSb without reporting the doping concentration; and Uribe~\textit{et al.}~\cite{munoz_uribe_near-band-gap_1996} reported the doping concentration but did not mention the growth method. 
(iii) Most experimental studies rely on SE to determine the refractive index, and SE comes with its own limitations. SE measures polarization changes upon reflection of a sample, typically thin-films on a substrate. As a consequence, samples with depolarization effects such as a rough surface or a non-uniform layer thickness, introduce errors. SE also requires an extensive data analysis, offers low spatial resolution, and has difficulties determining low absorption coefficients ($\alpha$ < \SI{100}{\per\centi\metre}, with $\alpha \propto k$)~\cite{fujiwara_spectroscopic_2009}. Moreover, the layer thickness is extracted indirectly from optical interference within the thin-film in a transparent wavelength region, hence it is inherently dependent on the optical measurement. (iv) Most studies do not report uncertainty estimates for their results, therefore a quantitative assessment and comparison are not possible, which makes the reported values unsuitable for device design. 

In this study, we present an alternative approach to refractive index measurements based on Perner~\textit{et al.}~\cite{perner_simultaneous_2023}, that addresses the shortcomings above. We provide a complete description of sample preparation and avoid SE specific drawbacks, such as the indirect measurement of the layer thicknesses. Consequently, we give a quantitative uncertainty analysis of our results. Specifically, we measure the photometric transmission of a GaSb/AlAsSb heteroepitaxial structure and determine the individual layer thicknesses using a scanning electron microscope (SEM), in this way avoiding the indirect method used in SE. We then perform a nonlinear regression based on our measurements to fit a parameterized refractive index model for $n^*_{\text{GaSb}}$ and a Sellmeier equation for $n_{\text{AlAsSb}}$. Finally, we quantify our results with a Monte Carlo-based uncertainty estimation.

\section{Theoretical Background}\label{sec:2.Theory}
The electric permittivity of any material is given by
\begin{equation}\label{equ:dielectric}
    \epsilon(E) = \epsilon_1(E) + i  \epsilon_2(E),
\end{equation}
where $\epsilon_1(E)$ is the real part, $\epsilon_2(E)$ is the imaginary part of $\epsilon(E)$, and $E$ is the photon energy. The refractive index $n^*(E)$ of the material is related to $\epsilon(E)$ via
\begin{equation}\label{equ:refractive_dielectric}
    n^*(E) = \sqrt{\epsilon(E)}.
\end{equation}
This implies that $n^*(E) \in \mathbb{C}$, with
\begin{equation}\label{equ:refractive}
    n^*(E) = n(E) +ik(E),
\end{equation}
where $n(E)$ represents the real refractive index and $k(E)$ the extinction coefficient. From \cref{equ:dielectric,equ:refractive_dielectric,equ:refractive} we derive that $n(E)$ and $k(E)$ are functions of $\epsilon_1(E)$ and $\epsilon_2(E)$
\begin{equation}
    n(E) = \sqrt{\frac{\sqrt{\epsilon_1^2(E) +  \epsilon_2^2(E)} + \epsilon_1(E)}{2}},
\end{equation}
\begin{equation}
    k(E) = \sqrt{\frac{\sqrt{\epsilon_1^2(E) +  \epsilon_2^2(E)} - \epsilon_1(E)}{2}}.
\end{equation}
Thus, knowing $\epsilon(E)$, one can deduce the refractive index and extinction coefficient.  

\section{Methods}\label{sec:3. Methods}
\begin{figure}[htb]
    \centering
    \includegraphics[
      width=\linewidth,
    ]{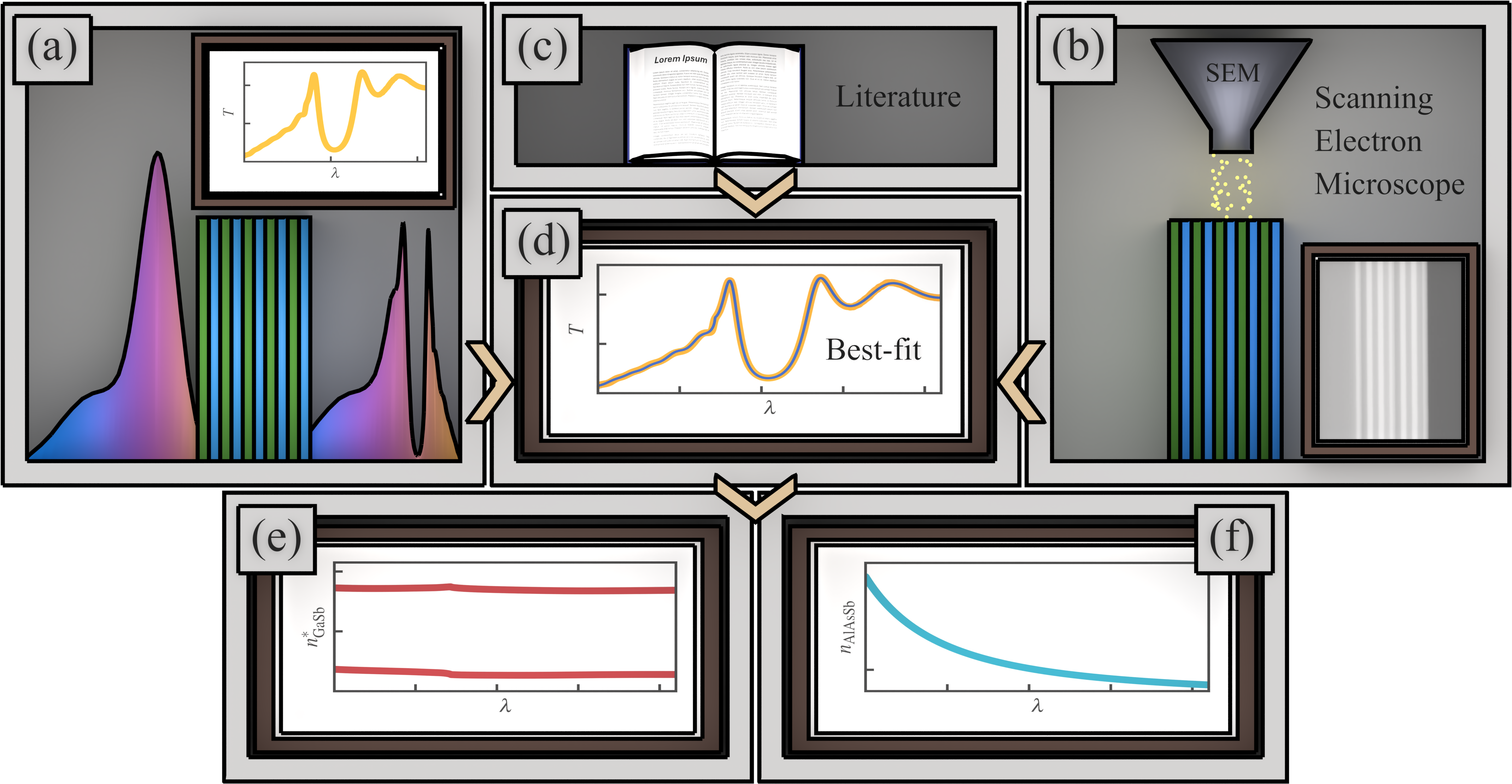}
    \caption{Overview of the workflow. (a) We measure the photometric transmission $T$ and (b) record cross-sectional images of the structure to extract the layer thicknesses \{$d_{\text{i}}$\}. (c) With literature values for $n_{\text{Sub}}$, we (d) perform a nonlinear fit of a theoretical model for $n^*_\text{GaSb}$ and the Sellmeier coefficients of $n_\text{AlAsSb}$ to $T$ with \{$d_{\text{i}}$\} as additional input. We use the fit parameters to calculate (e) $n^*_\text{GaSb}$ and (f) $n_\text{AlAsSb}$.} 
    \label{fig:concept}
\end{figure}

Fig.~\ref{fig:concept} illustrates our approach to determine $n^*_{\text{GaSb}}$ and $n_{\text{AlAsSb}}$.
As shown in Fig.~\ref{fig:concept}(a), we first measure the photometric transmission $T$ and subsequently, determine the layer thicknesses \{$d_{\text{i}}$\} from cross-sectional images of the heteroepitaxial structure (Fig.~\ref{fig:concept}(b)). We then fit a theoretical model for the complex refractive index of GaSb ($n^*_{\text{GaSb}}$) and a Sellmeier equation for the refractive index of AlAsSb ($n_{\text{AlAsSb}}$) to $T$ (Fig.~\ref{fig:concept}(d)), using the measured \{$d_{\text{i}}$\} as additional input. For the refractive index of the substrate ($n_{\text{Sub}}$), we use literature values (Fig.~\ref{fig:concept}(c)). Finally, we determine $n^*_{\text{GaSb}}$ and $n_{\text{AlAsSb}}$ from the fit parameters (Fig.~\ref{fig:concept}(e) and (f)).

The measurements of $T$ and the cross-sectional images (Fig.~\ref{fig:concept}(a) and (b)) are described in detail in Sec.~\ref{subsec:4.Transmission} and~\ref{subsec:5.CrossSection}, respectively. The determination of \{$d_{\text{i}}$\} is described in Sec.~\ref{subsec:6.LayerExtraction}.

To perform a nonlinear fit to $T$ (Fig.~\ref{fig:concept}(d)), we first require a mathematical framework. 
The dependence of $T$ on $n^*_{\text{GaSb}}$, $n_{\text{AlAsSb}}$, and $n_{\text{Sub}}$ is fully governed by the Fresnel’s equations and the respective layer thicknesses $\{d_{\text{i}}\}$. Thus, we model $T$ using the transfer-matrix method (\text{TMM})
\begin{equation}\label{equ:tmm}
    T = \text{TMM}(n^*_{\text{GaSb}},n_{\text{AlAsSb}},\{d_{\text{i}}\},n_{\text{Sub}}).
\end{equation}
This equation also highlights the importance of independently measuring $\{d_\text{i}\}$. The quantity $n^*_{\text{GaSb}}$ can be written as a function of the photon energy $E$ and the material-specific parameters, (the fit parameters $\Vec{\theta}$), as
\begin{equation}
    n^*_{\text{GaSb}} = n^*_{\text{GaSb}}(E, \Vec{\theta}\,).
\end{equation}
The full theoretical model for $n^*_{\text{GaSb}}$ is explained in detail in Sec.~\ref{subsec:2.refractive index mode}. The refractive index $n_{\text{AlAsSb}}$ is described by a Sellmeier equation\footnote{The Sellmeier equation is based on unpublished experimental measurements communicated privately by Dr.~V.~J.~Wittwer from the Université de Neuchâtel.}
\begin{equation}\label{equ:sellmeier}
        n_{\text{AlAsSb}}^2 - 1 = \frac{B\lambda^2}{\lambda^2 - B_0} + \frac{C\lambda^2}{\lambda^2 - C_0}, 
\end{equation}
with $\lambda$ denoting the wavelength and $B$, $B_0$, $C$, and $C_0$ being Sellmeier coefficients. The value of $n_{\text{Sub}}$ is taken from Aspnes~\textit{et al.}~\cite{skauli_improved_2003}.
We can now rewrite Eq.~\ref{equ:tmm} as
\begin{equation}\label{equ:tmm_new}
    T = \text{TMM}(E,\Vec{\theta}\,,B,B_0,C,C_0,\{d_{\text{i}}\},n_{\text{Sub}}),
\end{equation}
showing the explicit dependence of the model parameters and Sellmeier coefficients.
Fig.~\ref{fig:concept}(d) shows the nonlinear fit procedure and highlights the various input parameters. We fit Eq.~\ref{equ:tmm_new} to the measured $T$, using the measured \{$d_{\text{i}}$\} as additional inputs, and extract the model parameters $\Vec{\theta}$ and the Sellmeier coefficients. The full fit procedure is described in Sec.~\ref{subsec:7.FitProcedure}.
In the last step, depicted in Fig.~\ref{fig:concept}(e) and (f), we use these fitted parameters to simultaneously calculate $n^*_{\text{GaSb}}$ and $n_{\text{AlAsSb}}$.

\subsection{Refractive Index Model for GaSb}\label{subsec:2.refractive index mode}
To fit Eq.~\ref{equ:tmm_new} we need a theoretical model for $n^*_{\text{GaSb}}$. For GaSb, several different such models exist. However, most of them are either to intricate to be used in a fit procedure (see Ref.~\cite{kim_modeling_1992,paskov_refractive_1997, linnik_calculations_2002}) or are inaccurate in their description (see Ref.~\cite{adachi_optical_1989}). We use the model of Djuri\v{s}i\'c \textit{et al.}~\cite{djurisic_modeling_2000} because it yields good results, offers a comprehensive description, and is straightforward to implement.

The following section is based on the work of Djuri\v{s}i\'c \textit{et al.}~\cite{djurisic_modeling_2000}. For a detailed discussion of the individual equations and parameters, we refer to the original literature. Here, we present only the formulas used. 

The authors separate the electric permittivity into a sum of its constituents, 
\begin{equation}
    \epsilon(E) = \epsilon_{1\infty} + \epsilon_\text{I}(E) + \epsilon_\text{II}(E) + \epsilon_\text{III}(E),
\end{equation}
where each contribution corresponds to different critical points (CPs) within the material's Brillouin zone. In total, the model of Djuri\v{s}i\'c \textit{et al.} has 24 parameters. 

\paragraph{\bm{$\epsilon_{1\infty}:$}}
$\epsilon_{1\infty}$ is a constant describing CP transitions outside the investigated spectral range~\cite{djurisic_modeling_2000}. 
 
\paragraph{\bm{$\epsilon_\text{I}(E):$}}
 The $E_0$ and $E_0 + \Delta_0$ CPs are described by
\begin{equation}\label{equ:epsilon1}
    \epsilon_\text{I}(E) = AE_0^{-3/2} \left[ f(\chi_0) + \frac{1}{2}\left( \frac{E_0}{E_0 + \Delta_0}  \right)^{3/2} f(\chi_{0\text{s}})  \right],
\end{equation}
where $A$ is the strength parameter for this transition and $f(\chi_0)$ and $f(\chi_{0\text{s}})$ are given by 
\begin{equation}\label{equ:f}
    f(y) = y^{-2} \left[ 2 - (1+y)^{1/2} - (1-y)^{1/2}      \right].
\end{equation}
$\chi_0$ and $\chi_{0\text{s}}$ are defined via
\begin{align}\label{equ:chis}
    \chi_n = \frac{E + i \Gamma_n}{E_n}, & \qquad \chi_{n\text{s}} = \frac{E + i\Gamma_n}{E_n + \Delta_n},
\end{align}
with $n=0$.
$\Gamma_0$ is the damping parameter for the $E_0$ and $E_0 + \Delta_0$ transition.

\paragraph{\bm{$\epsilon_\text{II}(E):$}}
The $E_1$ and $E_1 + \Delta_1$ CP transitions are expressed through
\begin{equation}\label{equ:epsilon2}
    \epsilon_\text{II} = -B_1 \chi_1^{-2} \ln(1-\chi_1^2) - B_{1\text{s}} \chi_{1\text{s}}^{-2} \ln(1-\chi_{1\text{s}}^2),
\end{equation}
where $\chi_1$ and $\chi_{1\text{s}}$ are obtained by Eq.~\ref{equ:chis} with $n=1$. $B_1$ and $B_{1\text{s}}$ are strength parameters for this transition and $\Gamma_1$ is the damping coefficient.

\paragraph{\bm{$\epsilon_\text{III}(E):$}}
Higher lying CP transitions are modeled by damped harmonic oscillators
\begin{equation}
    \epsilon_\text{III} = \sum_{j=2}^{4} \frac{F^2_j}{E_j^2 - E^2 - iE\Gamma_j}
\end{equation}
with strength $F_j = \sqrt{f_jE_j^2}$. 

For each transition, the authors introduce a frequency-dependent damping function
\begin{equation}\label{equ:freq}
    \Gamma_j(E) = \Gamma_j \exp\left[- \alpha_j \left(\frac{E - E_j}{\Gamma_j}\right)^2\right],
\end{equation}
where $\Gamma_j$ are damping constants, $E_j$ are transition energies, and $E$ is the photon energy. $\alpha_j$ are parameters to continuously change the lineshapes. According to the authors, Eq.~\ref{equ:freq} is the main driver for improvements in accuracy compared to other models.

The complex refractive index $n^*(E)$ is computed via Eq.~\ref{equ:refractive_dielectric}.

\subsection{Sample Preparation}\label{subsec:3.SamplePreparation}
The sample characterized in this work is based on an epitaxial heterostructure designed as a distributed Bragg reflector (DBR) for a center wavelength of \SI{2.05}{\micro\meter}. We grow the epitaxial heterostructure using a Veeco Gen III research MBE reactor. The growth is performed on a tellurium-doped, \SI{500}{\micro\meter}-thick, 2-inch single-side polished GaSb substrate supplied by Wafer Technology Ltd. The layer sequence starts by depositing a \SI{250}{\nm} thick $\mathrm{InAs_{0.91}Sb_{0.09}}$ layer that serves as an etch-stop to facilitate subsequent substrate removal. This layer is then capped by the 6.5-period GaSb/$\mathrm{AlAs_{0.08}Sb_{0.92}}$ DBR. The resulting heteroepitaxial structure has a total thickness of \SI{2.14}{\micro\meter}, with the DBR investigated in this work constituting \SI{1.9}{\micro\meter}. Further details regarding the growth parameters can be found in \cite{gaulke_high_2021}.

Following growth, the wafer is cleaved into $1.5 \times \SI{1.5}{\cm^2}$ samples. Larger ($1.6 \times \SI{1.6}{\cm^2}$) double-side polished, undoped GaAs pieces are prepared for use as bonding carriers.
GaAs is chosen for the bonding carriers due to its transparency above \SI{808}{\nano \metre}. Both, the samples and carrier pieces, underwent a solvent cleaning procedure (methanol, isopropyl alcohol, and acetone) as described in Ref.~\cite{schuchter_2-m_2024}, followed by surface activation via oxygen plasma (Technics Plasma 100-E). Direct wafer bonding is subsequently performed in a wafer bonder (Applied Microengineering Ltd), applying a force of \SI{1000}{\newton} at \SI{300}{\celsius}. The GaSb substrate is then removed in a two-stage process. First, the majority of the GaSb substrate is mechanically thinned by lapping to a residual thickness of $\approx$\SI{20}{\micro\meter}. Second, the remaining GaSb substrate is chemically removed using a $\mathrm{CrO_{3}:HF: H_2O}$ wet-etching solution. Finally, the etch-stop layer is selectively removed using a citric acid and hydrogen peroxide solution. A comprehensive description of this fabrication workflow is available in Refs.~\cite{gaulke_high_2021, schuchter_2-m_2024, huwyler_3-w_2023}.

\subsection{Transmission Measurement}\label{subsec:4.Transmission}
\begin{figure}[htb]
    \centering
    \includegraphics[width=1\linewidth]{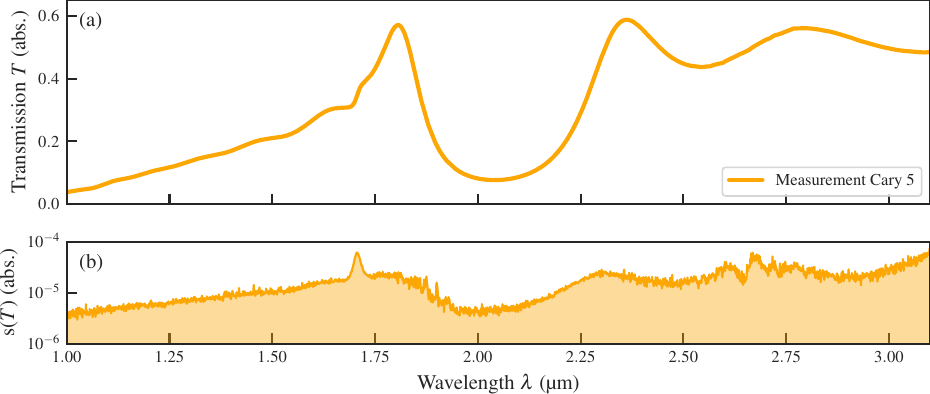}
    \caption{(a) The baseline-corrected transmission of the heteroepitaxial sample, measured using a Varian Cary 5 in the wavelength range from \SIrange{1}{3.1}{\micro \metre}. (b) 1$\sigma$ standard uncertainty of the mean measurement shown in panel (a).}
    \label{fig:transmission_measurement}
\end{figure}

We measure the photometric transmission of the sample with a Varian Cary 5 UV-Vis-NIR spectrophotometer at normal incidence. The spectrophotometer covers a wavelength range of \SIrange{0.175}{3.3}{\micro\metre} with an accuracy of \SI{\pm 0.1}{\nano\metre} in the UV-Vis spectral range and \SI{\pm 0.4}{\nano\metre} in the NIR spectral range. During the measurements, the Cary 5 operates in double beam mode and records spectra in the wavelength range from \SIrange{0.9}{3.3}{\micro\metre}. We measure at a scan rate of \SI{600}{\nano\metre \per \minute} with a resolution of \SI{1}{\nano\metre}, and an average time of \SI{0.1}{\second}. During this \SI{0.1}{\second} period, the Cary 5 averages three measured data points. To account for this, we apply a scaling factor of $3^{-1/2}$ to the 1$\sigma$ standard uncertainty of the mean ($s(T)$). 
We set the beam size with a \SI{7.5}{\milli\metre} precision pinhole (Thorlabs, P7500K) mounted to a custom made sample holder and measure three different types of spectra:
\begin{enumerate}[label=(\roman*), style=nextline]
    \item 0\% transmission spectrum ($T_{\text{0\%}}$), where the light path is blocked and we measure the dark counts of the detector.
    \item 100\% transmission spectrum ($T_{\text{100\%}}$), where we measure the full throughput without the sample.
    \item Sample transmission spectrum ($T_{\text{sample}}$), where the sample is placed in the light path of the Cary 5.
\end{enumerate}
The background spectra ($T_{\text{0\%}}$ and $T_{\text{100\%}}$) are measured ten times and $T_{\text{sample}}$ is measured 100 times. We calculate the average of each spectrum type and apply the baseline correction formula
\begin{equation}
\label{equ:transmission}
    T = \frac{T_{\text{sample}} - T_{\text{0\%}}}{T_{\text{100\%}} - T_{\text{0\%}}},
\end{equation}
to the averages to obtain the background corrected photometric transmission spectra of our sample ($T$).
We propagate $s(T)$ through Eq.~\ref{equ:transmission} using the \texttt{uncertainties} package for Python 3~\cite{lebigot_uncertainties_2024}.
Fig.~\ref{fig:transmission_measurement}(a) shows $T$, while Fig.~\ref{fig:transmission_measurement}(b) presents $s(T)$. Due to high detector noise at longer wavelengths (>\SI{3.1}{\micro\metre}) and high absorption of GaSb at shorter wavelengths (<\SI{1}{\micro \metre}), we restrict our data to the \SIrange{1}{3.1}{\micro\metre} spectral range. 
The entire measurement run (\textasciitilde \SI{9}{h}) is conducted in a climate-controlled laboratory, with the temperature maintained at \SI{20.4 \pm 0.5}{\degreeCelsius}. Water vapor absorption lines are removed during the baseline correction (Eq.~\ref{equ:transmission}). 

\subsection{Cross-Section Measurement}\label{subsec:5.CrossSection}
\begin{figure}[htb]
    \centering
    \includegraphics[width=1\linewidth]{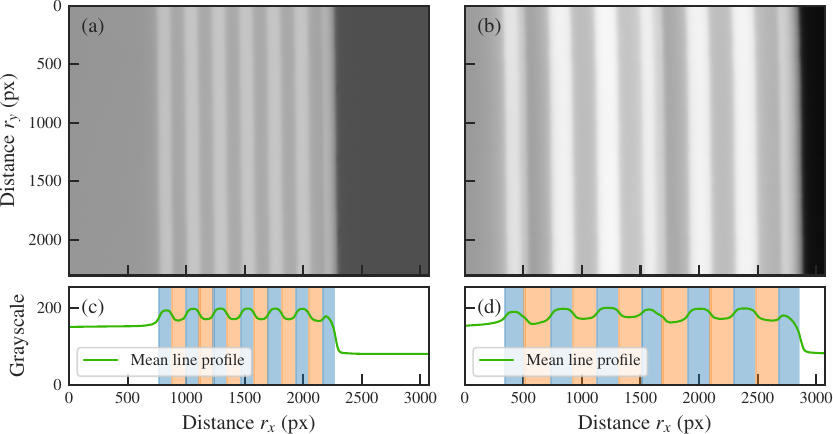}
    \caption{Representative cross-sectional grayscale image of the sample with (a) \SI{30}{\kilo X} and (b) \SI{50}{\kilo X} magnification, respectively. The green line shows the column-wise mean grayscale intensity for (c) \SI{30}{\kilo X} and (d) \SI{50}{\kilo X} magnification. The orange and blue stripes indicate the extracted layer thicknesses.}
    \label{fig:SEM}
\end{figure}

We conduct the cross-sectional measurements using a SEM (Zeiss Supra 55 VP) equipped with a four-quadrant backscattered electron detector. Before starting the measurements, we evacuate the SEM to a pressure of approximately \SI{2e-6}{\milli\bar}, set the working distance to \SI{9.4}{\milli\metre}, and the beam energy to \SI{10}{\kilo\electronvolt}. Afterwards, we acquire calibration images at \SI{30}{\kilo X} and \SI{50}{\kilo X} magnification using a NIST-certified calibration standard (EM-TEC MCS-0.1CF). We then record cross-sectional images of our sample. At each position, we take images at both \SI{30}{\kilo X} (Fig.~\ref{fig:SEM}(a)) and \SI{50}{\kilo X} (Fig.~\ref{fig:SEM}(b)) magnification. Fig.~\ref{fig:SEM}(c) and~\ref{fig:SEM}(d) show the column-wise calculated mean line profile (green line), with the orange and blue stripes marking the extracted layer thicknesses.

\subsection{Layer Thickness Extraction}\label{subsec:6.LayerExtraction}
To extract the layer thicknesses from the recorded grayscale images (see Fig.~\ref{fig:SEM}), we use a custom Python 3 program, similar to that described in Ref.~\cite{perner_simultaneous_2023}. First, we apply the program to the images of the calibration standard (see Sec.~\ref{subsec:5.CrossSection}) and determine a calibrated pixel pitch of
\SI{1.23622 \pm 0.00007}{\nano\metre \per \pixel} at \SI{30}{\kilo X} magnification and \SI{0.74267 \pm 0.00006}{\nano\metre\per\pixel} at \SI{50}{\kilo X} magnification, respectively. These calibration values are later used in \textbf{Step 5} to convert the layer thickness from pixels into a physical distance in \si{\micro\metre}.

The program's workflow is described below:
\begin{description}[style=nextline, leftmargin=!, labelwidth=\widthof{\bfseries Step 4:}]
    \item[Step 1:] 
    Load the image into a 2D array. We notice that the images have blank lines at the top and the bottom, hence we consider only rows $r_{\text{y}}$ from 100 to 2200 and columns $r_{\text{x}}$ from 10 to 3000.
    \item[Step 2:] Compute the column-wise mean intensity to obtain the average line profile (green line in Fig.~\ref{fig:SEM}(c),(d)). Differentiate this line profile and take the position of its local extrema as estimates for the interface position between neighboring layers.
    \item[Step 3:] Loop over all rows in the 2D array:
    \begin{enumerate}[label=(\roman*), style=nextline]
    \item  Split the row into symmetric intervals around the previously estimated interfaces, ensuring that each interval contains exactly one interface.
    \item Fit an error function to each interval. The gradual intensity transition observed at the layer interfaces is accurately modeled by an error function, where the inflection point of the fitted curve estimates the location of the interface.  
    \item The difference between adjacent inflection points corresponds to the layer thickness in pixels.
    \end{enumerate}
    \item[Step 4:] Calculate the mean and 1$\sigma$ standard uncertainty of the mean $s(d_{\text{i}})$ for each extracted layer thickness $d_{\text{i}}$ of the sample. 
    \item[Step 5:] \label{num:calibrate} Multiply by the calibrated pixel pitch to convert $\{d_{\text{i}}\}$ into physical distances in \si{\micro\metre}.
    \item[Step 6:]  Combine the mean values and 1$\sigma$ standard uncertainty of the mean $s(\{d_{\text{i}}\})$ of both magnifications. We use a weighted average to account for the different $s(\{d_{\text{i}}\})$.
\end{description}

\begin{figure}[htb]
    \centering
    \includegraphics[width=1\linewidth]{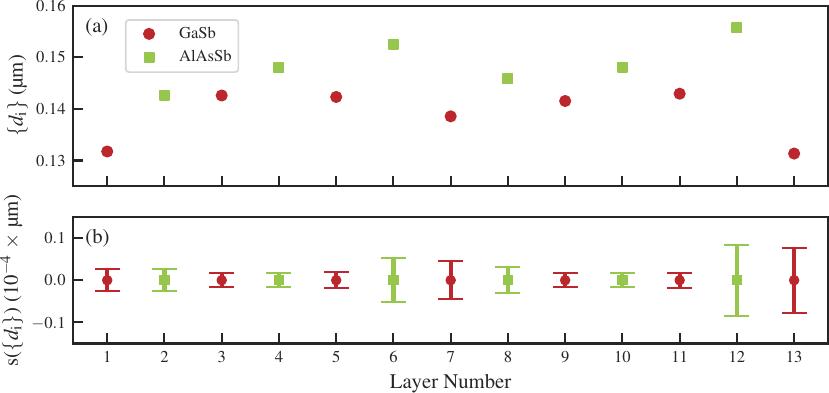}
    \caption{(a) The extracted physical layer thicknesses $\{d_{\text{i}}\}$. Red circles corresponds to GaSb and green squares to AlAsSb, respectively. (b) The 1$\sigma$ standard uncertainty $s(\{d_{\text{i}}\})$ of the individual layer thicknesses. Note the scaling factor $10^{-4}$ in front.}
    \label{fig:Layers}
\end{figure}

The extracted mean layer thicknesses $\{d_{\text{i}}\}$ are displayed in Fig.~\ref{fig:Layers}(a). The corresponding 1$\sigma$ standard uncertainty $s(\{d_{\text{i}}\})$ is shown in Fig.~\ref{fig:Layers}(b).

\subsection{Nonlinear Fit Procedure}\label{subsec:7.FitProcedure}

The fitting framework is built on the Python 3 library \texttt{lmfit}~\cite{newville_lmfit_2025} and the best fit is found by minimizing the residuals $r_i$ 
\begin{equation}\label{equ:residuals}
    r_i = T - T_{\text{model}},
\end{equation}
where $T$ is described in Sec.~\ref{subsec:4.Transmission} and $T_{\text{model}}$ denotes the simulated transmission calculated using the \texttt{tmm-fast} library for Python 3~\cite{luce_tmm-fast_2022}.
To perform this minimization, the Log-Cosh Loss objective function $L(r_i)$ 
\begin{equation}
    L(r_i) = \sum_{i=1}^{n} \log(\cosh(r_i)),
\end{equation}
is used since it is smooth (unlike the mean absolute error), less sensitive to outliers than the mean squared error, and does not need an additional parameter such as the Huber loss~\cite{terven_comprehensive_2025}.

As mentioned in Sec.~\ref{sec:3. Methods} (see also Fig.~\ref{fig:concept}(d)), we fit Eq.~\ref{equ:tmm_new}, with the measured layer thicknesses $\{d_\text{i}\}$ as input, to the photometric transmission $T$. The fit parameters correspond to the Sellmeier coefficients and the model parameters $\Vec{\theta}$ as described in Sec.~\ref{sec:3. Methods} and~\ref{subsec:2.refractive index mode}.

The nonlinear fit procedure is structured as follows: (i) we first identify parameters (that are Sellmeier coefficients and $\Vec{\theta}$) by finding a global optimum using a single run of a time-expensive global optimizer, (ii) we refine these found parameters with a fast local optimizer. We use this approach, because initial parameters are crucial for finding optimal solutions for our non-convex objective function, and this way we can start the second optimization within a convex region near the global optimum. The global optimizer we use is Adaptive Memory Programming for Constrained Global Optimization (AMPGO)~\cite{lasdon_adaptive_2010} and the local optimizer is Nelder-Mead simplex algorithm~\cite{nelder_simplex_1965,gao_implementing_2012}. 
 
\begin{figure}[htb]
    \centering
    \includegraphics[width=1\linewidth]{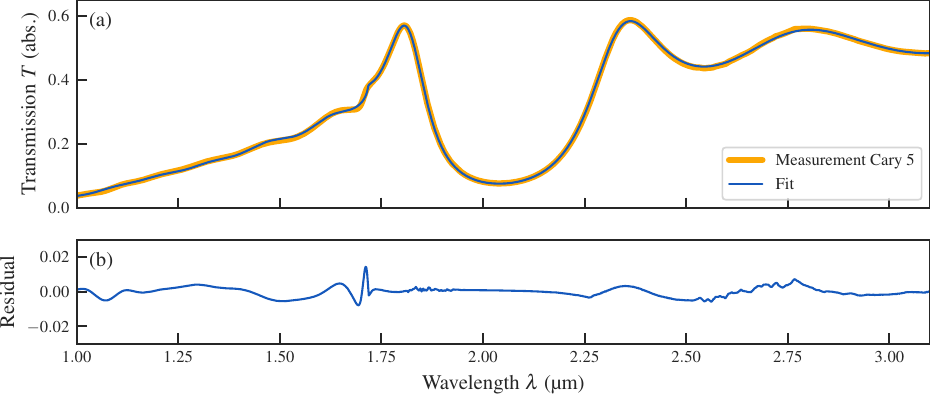}
    \caption{(a) The measured photometric transmission $T$ (thick yellow line) together with the best-fit result (thin blue line). (b) The corresponding residual, obtained via Eq.~\ref{equ:residuals}.}
    \label{fig:all_together}
\end{figure}

The fit result is presented in Fig.~\ref{fig:all_together}(a) (blue line). The relative root mean square error (RRMSE) is \SI{0.75}{\percent}, suggesting strong agreement between measurement and fit. This is especially evident from the residuals depicted in Fig.~\ref{fig:all_together}(b). The sharp edge at $\approx$\SI{1.7}{\micro \metre} is inherent to the refractive index model of GaSb and can be traced back to Eq.~\ref{equ:f}, in particular the term $\sqrt{1-y}$, where $y$ is a complex number.

\subsection{Uncertainty Estimation}\label{subsec:8.Error}
The uncertainties in $n^*_{\text{GaSb}}$ and $n_{\text{AlAsSb}}$ are evaluated using Eq.~\ref{equ:tmm_new}. However, a straightforward analytical uncertainty propagation is not available because TMM is an intricate function and the model parameters are obtained via a nonlinear fit~\cite{perner_simultaneous_2023}. Consequently, we employ a numerical (Monte Carlo) approach to propagate the uncertainties, following Ref.~\cite{perner_simultaneous_2023}, which allow us to propagate the known uncertainties of the experimental inputs through the fitting procedure. 
 
The quantities in Eq.~\ref{equ:tmm} that carry uncertainties are the measured layer thicknesses \{$d_\text{i}$\}, the measured transmission $T$, and the refractive index of the substrate $n_{\text{Sub}}$. To account for these uncertainties, we execute the nonlinear fit procedure 500 times (only the fast local optimizer), each time with a different realization of said quantities. Each fit yields distinct Sellmeier coefficients and model parameters $\Vec{\theta}$, as a result giving different values for $n^*_{\text{GaSb}}$ and $n_{\text{AlAsSb}}$. Finally, we determine the 1$\sigma$ standard uncertainty of the mean for $n^*_{\text{GaSb}}$ and $n_{\text{AlAsSb}}$, which represents the uncertainty in the measured refractive indices.

We achieve the variation in $T$, \{$d_{\text{i}}$\}, and $n_{\text{Sub}}$ with a resampling approach. For that, we randomly selected the values from a normal distribution centered at the mean values with corresponding standard deviation. For the values used in $T$ and \{$d_{\text{i}}$\}, see Figs.\ref{fig:transmission_measurement} and \ref{fig:Layers}, respectively. For $n_{\text{Sub}}$, we use an uncertainty of \SI{0.2}{\percent}, following Ref.\cite{skauli_improved_2003}.

\section{Results and Discussion}\label{sec:4.Result}

\begin{figure}[htb]
    \centering
    \includegraphics[width=1\linewidth]{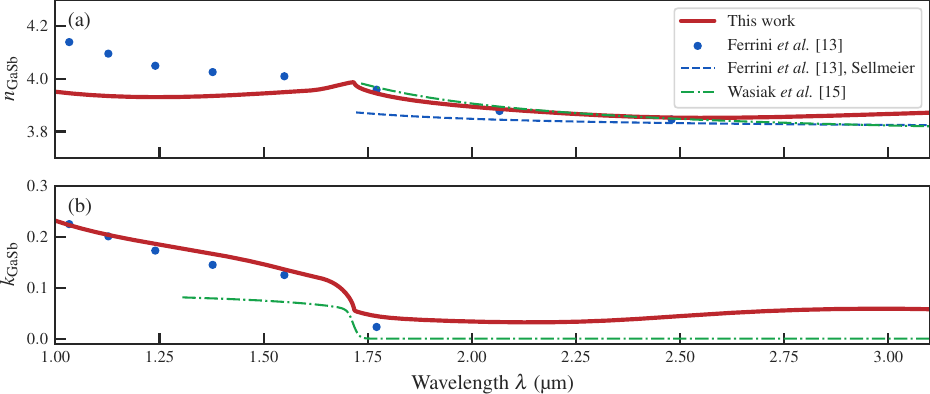}
    \caption{Our results for $n^*_{\text{GaSb}}$, shown as solid red lines: (a) The real part $n_{\text{GaSb}}$, (b) the imaginary part $k_{\text{GaSb}}$. For comparison, we include experimental values from Ref.~\cite{ferrini_optical_1998} (blue dots and blue dashed line) and Ref.~\cite{wasiak_absorption_2018} (green dot-dashed line). Note: Ref.\cite{wasiak_absorption_2018} reports a range of values for the \textit{A} parameter of the dispersion relation. The values shown here correspond to \textit{A} = 12.45.}
    \label{fig:refractive_index_ngasb}
\end{figure}

The real and imaginary parts of $n^*_{\text{GaSb}}$ in the spectral range \SIrange{1}{3.1}{\micro \metre} are shown in Fig.~\ref{fig:refractive_index_ngasb}(a) and (b), respectively, as solid red lines. The fit results of the model parameters $\Vec{\theta}$ are summarized in Tab.~\ref{tab:fit_results_ngasb}. We compare our data with the results of Ferrini \textit{et al.}~\cite{ferrini_optical_1998}, which were used to derive the refractive index model employed in this work, and with the most recent values found, reported by Wasiak \textit{et al.}~\cite{wasiak_absorption_2018}.

\begin{table}[!ht]
\centering

\begin{tabular}{*{8}{c}}
\toprule
$E_0$ (\si{\eV})& $E_1$ (\si{\eV})& $E_2$ (\si{\eV}) & $E_3$ (\si{\eV})& $E_4$ (\si{\eV})   &  $\Delta_0$ (\si{\eV}) & $\Delta_1$ (\si{\eV}) & $\epsilon_{1\infty}$ \\ 
\midrule
\num{0.722} & \num{2.862} & \num{0.739} & \num{0.421} & \num{3.797}&  \num{1.092} & \num{0.020} &\num{8.427e-07}  \\
\bottomrule
\end{tabular}

\vspace{0.1cm}

\begin{tabular}{*{8}{c}}
\toprule
 $\Gamma_0$ (\si{\eV})  & $\Gamma_1$ (\si{\eV})  & $\Gamma_2$ (\si{\eV}) & $\Gamma_3$ (\si{\eV}) & $\Gamma_4$ (\si{\eV}) & $F_2$ (\si{\eV})& $F_3$ (\si{\eV})& $F_4$ (\si{\eV})  \\ 
\midrule
 \num{3.213e-4} & \num{0.949} & \num{0.053} & \num{0.172} & \num{0.095}  & \num{0.067} & \num{0.153} & \num{4.256} \\ 
\bottomrule
\end{tabular}

\vspace{0.1cm}

\begin{tabular}{*{8}{c}}
\toprule
$\alpha_0$ & $\alpha_1$ & $\alpha_2$ & $\alpha_3$ & $\alpha_4$ &   $A$ (\si{\eV^{3/2}}) & $B_1$ & $B_{1s}$   \\ 
\midrule
\num{2.463e-4} & \num{0.239} & \num{0.933} & \num{1.602} & \num{0.013} & \num{1.488} & \num{12.770} & \num{1.512e-4} \\
\bottomrule
\end{tabular}

\vspace{5pt}

\caption{Our fit results for the model parameters $\Vec{\theta}$ of $n^*_{\text{GaSb}}$. The values are rounded to three decimal points.}
\label{tab:fit_results_ngasb}
\end{table}

\paragraph{Real part of  $n^*_{\text{GaSb}}$:}
The available data shows that all three curves overlap in the \SIrange{1.75}{2.50}{\micro \metre} wavelength region. Above \SI{2.50}{\micro \metre} Wasiak’s results and ours slightly deviate with a maximum of $\approx$0.052 at \SI{3.1}{\micro \metre}. This deviation arises from the fact that Wasiak used a strictly monotonically decreasing Sellmeier equation over the full reported wavelength range ($\approx$\SIrange{1.73}{11.8}{\micro \metre}). The part of Ferrini’s data which are based on a Sellmeier equation (dashed blue line) do not agree with our results, but align, at longer wavelengths, with those of Wasiak. Below $\approx$\SI{1.75}{\micro \metre}, Wasiak provides no data, and our results diverge from Ferrini’s (blue dots), which were obtained by methods other than the Sellmeier equation used at longer wavelengths. Our work shows a relatively flat refractive index behaviour in that region, whereas Ferrini's data presents an ascending trend with decreasing wavelength. The reason for this may lie in their methodology: first, they do not directly measure the layer thickness; instead, they infer it from the amplitude and period of interferene oscillations in the measured $R$ spectrum. Second, the refractive index in that region was determined from separate measurements (interference fringes, SE, and KK of $R$). This is also supported by the fact that the different wavelength ranges reported by Ferrini do not connect at the band gap wavelength (blue dots and blue dashed line). 
Finally, we note the peak near the GaSb band gap at $\approx$\SI{0.72}{\eV} ($\approx$\SI{1.7}{\micro\metre}), the same characteristic is reported in Refs.~\cite{paskov_refractive_1997, alibert_refractive_1991}. 

\paragraph{Imaginary part of  $n^*_{\text{GaSb}}$:}
Unlike the real part of the refractive index, no spectral region shows perfect overlap among all three curves. Ferrini's data and ours agree reasonably well across $\approx$\SIrange{1}{1.75}{\micro \metre}, while Wasiak's values are consistently lower. Note, that Wasiak report the absorption coefficient $\alpha$ for \SIrange{0.28}{0.95}{\eV} ($\approx$\SIrange{4.43}{1.3}{\micro \metre}) instead of the extinction coefficient $k_{\text{GaSb}}$. Accordingly, the values for $k_{\text{GaSb}}$ in Fig.~\ref{fig:refractive_index_ngasb}(b) labeled "Wasiak~\textit{et al.} [15]" are computed using $k = \frac{c h \alpha}{E 4 \pi}$, with $c$ the speed of light, $h$ Planck's constant. Both datasets, Wasiak's and ours, do capture the steep decrease in $k_{\text{GaSb}}$ around \SI{1.75}{\micro \metre} and Wasiak attribute it to the Urbach tail.
Wasiak \textit{et al.} reports absorption near \SI{0.4}{\eV} ($\approx$\SI{3.1}{\micro \metre}), which they attributed to residual absorption of the substrate; in contrast, we observe an increasing trend of $k_{\text{GaSb}}$ at longer wavelengths. Ferrini does not report values for $k_{\text{GaSb}}$ for wavelengths higher than $\approx$\SI{1.75}{\micro \metre}.
Below $\approx$\SI{1.75}{\micro \metre}, both our data and Ferrini's exhibit an ascending trend, whereas Wasiak's shows minimal or no increase. 

In summary, our results for $n^*_{\text{GaSb}}$ in the wavelength range from \SIrange{1}{3.1}{\micro \metre} show deviations from literature values in specific regions, which we attribute to the differeing approaches and analysis used in the various works. As noted above, Ferrini~\textit{et al.} rely on an indirect measurement of the layer thickness, and connected multiple methods to obtain the refractive index, including a numerical KK transform based on an extrapolated $R$ spectrum.  While these calculations are justified, they introduce a larger margin of error, evident from the not ideal continuity between the Sellmeier-based values (Fig.~\ref{fig:refractive_index_ngasb}(a) dashed blue line) and the points obtained by the other methods (Fig.~\ref{fig:refractive_index_ngasb}(a) blue dots). By contrast, Wasiak \textit{et al.} determined the layer thickness by SEM. They derived $n_{\text{GaSb}}$ by fitting a Sellmeier model to interference fringes in transmission spectra over the entire reported wavelength range, and they report $\alpha$ rather than $k_{\text{GaSb}}$. 
In our work, we determine the refractive index using a single measurement technique (photometric transmission) across the entire reported wavelength range, measure the layer thicknesses independently of the optical data using SEM, and employ a dedicated theoretical model for $n^*_{\text{GaSb}}$ that describes the real and imaginary parts within a single framework.

\begin{figure}[htb]
    \centering
    \includegraphics[width=1\linewidth]{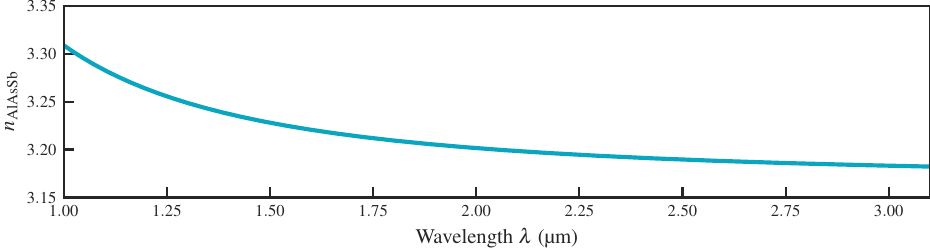}
    \caption{Our result for the refractive index of $\mathrm{AlAs_{0.08}Sb_{0.92}}$ using a Sellmeier equation.}
    \label{fig:refractive_index_nalassb}
\end{figure}

The result for $n_{\text{AlAsSb}}$ is shown in Fig.~\ref{fig:refractive_index_nalassb} and the fitted Sellmeier coefficients are summarized in Tab.~\ref{tab:fit_results_alassb}. To the best of our knowledge, broadband values for $n_{\text{AlAsSb}}$ with the same mole fraction as in our sample (x = 0.08) are not available in the literature.

\begin{table}[!ht]
\centering
\begin{tabular}{*{4}{c}}

\toprule

$B$ & $B_0$ (\si{\micro \metre ^2})  & $C$ & $C_0$ (\si{\micro \metre ^2})  \\ 
\midrule
\num{1.545} & \num{0.017} & \num{7.496} & \num{0.105} \\ 
\bottomrule
\end{tabular}

\vspace{5pt}
\caption{Fitted Sellmeier coefficient (Eq.~\ref{equ:sellmeier}) for AlAsSb, rounded to three decimal points.}
\label{tab:fit_results_alassb}
\end{table}

The uncertainty estimation described in Sec.\ref{subsec:8.Error}, applied to our results, yield the relative uncertainties presented in Tab.~\ref{tab:uncertainties}.

\begin{table}[!ht]
\centering
\begin{tabular}{*{3}{c}}

\toprule

$s(n_\text{GaSb})/n_\text{GaSb}$ & $s(k_\text{GaSb})/k_\text{GaSb}$  & $s(n_\text{AlAsSb})/n_\text{AlAsSb}$  \\ 
\midrule
\num{7.8e-5} & \num{2.0e-3} & \num{3.9e-4} \\ 
\bottomrule
\end{tabular}

\vspace{5pt}
\caption{Relative uncertainties for $n^*_\text{GaSb} = n_\text{GaSb} + ik_\text{GaSb} $ and $n_\text{AlAsSb}$.}
\label{tab:uncertainties}
\end{table}

\section{Conclusion}\label{sec:5.Conclusion}
We report complex refractive index values for the III--V semiconductor GaSb and refractive index values for $\mathrm{AlAs_{0.08}Sb_{0.92}}$, measured in a controlled environment from \SIrange{1}{3.1}{\micro \metre}. The relative uncertainty are <\num{7.8e-5} for $n_{\text{GaSb}}$, <\num{2.0e-3} for $k_{\text{GaSb}}$, and <\num{3.9e-4} for $n_{\text{AlAsSb}}$. Measurements are performed on a \SI{6.5}{period}, MBE grown GaSb/AlAsSb thin-film heteroepitaxial structure. Our methodological approach combines measurements of the photometric transmission $T$ with calibrated cross-sectional SEM measurements of the individual layer thicknesses $\{d_{\text{i}} \}$ of the heterostructure. From these measurements, we simultaneously determine $n^*_{\text{GaSb}}$ and $n_{\text{AlAsSb}}$ by a nonlinear fit procedure. Additionally, we provide an uncertainty estimation based on a Monte Carlo approach to quantify our results. As stated in Ref.~\cite{perner_simultaneous_2023}, this approach can be applied to crystalline or amorphous multilayers and is not limited to DBRs, as it only requires that the sample's optical response shows broadband features unaltered by the spectrometer. Additionally, this method does not rely on specific equipment such as SE~\cite{perner_simultaneous_2023}.

We compare our results with literature values (Refs.~\cite{ferrini_optical_1998,wasiak_absorption_2018}) and find deviations in certain wavelength regions, which we attribute to differing methodological approaches. Ref.~\cite{ferrini_optical_1998} and Ref.~\cite{wasiak_absorption_2018} both use a Sellmeier equation for wavelengths longer than $\approx$\SI{1.75}{\micro\metre}, corresponding to the GaSb band gap. However, Sellmeier equations are derived from approximations to the Lorentz oscillator model and are valid only far from resonances~\cite{saleh_fundamentals_2018}. In contrast, we use a dedicated theoretical model (see Sec.~\ref{subsec:2.refractive index mode} and Ref.~\cite{djurisic_modeling_2000}) for $n^*_{\text{GaSb}}$. 
The layers thickness in Ref.~\cite{ferrini_optical_1998} was determined indirectly from interference fringes, which inherently ties the thickness to the optical measurements. In contrast, we measure the thicknesses independently using SEM, thereby decoupling the optical and the thickness measurements. Further, Ref.~\cite{ferrini_optical_1998} combines separate measurement approaches to determine $n^*_{\text{GaSb}}$, including interference fringes, SE, and KK calculations of the measured reflection spectrum $R$, while Ref.~\cite{wasiak_absorption_2018} and our work employ a single optical measurement technique across the entire reported wavelength range. This highlights the methodological differences among the works and helps explain the discrepancies in the reported results.

Finally, we conclude that our updated complex refractive index data for GaSb will support the design, development, and optimization of future passive and active optoelectronic devices.

\noindent\textbf{Data and Code Availability:}
The authors can provide data and code upon reasonable request.

\noindent\textbf{Acknowledgment:}
This research was funded in whole or in part by the Austrian Science Fund (FWF)[10.55776/F1004] and [10.55776/P36040]. For open access purposes, the author has applied a CC BY public copyright license to any author accepted manuscript version arising from this submission. We acknowledge support by Stephan Puchegger and Martin Haßler and the Faculty Center for Nano Structure Research at the University of Vienna. All sample preparation steps were conducted at the FIRST cleanroom facility at ETH Zürich. Instrument identification in this paper is for experimental clarity only and does not imply endorsement. 

\noindent\textbf{Author Contribution:} 
U.G.: Writing – original draft (lead), Writing – review and editing (lead), Visualization (lead), Project administration (lead), Data curation (lead), Investigation (lead), Formal analysis (lead), Software (lead).
N.H.: Resources (equal), Writing – original draft (supporting).
M.E.: Writing – review and editing (supporting).
M.G.: Resources (equal).
O.H.H.: Writing – review and editing (supporting), Supervision (lead), Conceptualization (lead).

\noindent\textbf{Declaration of Interests:}
The authors declare no conflicts of interest.

\noindent\textbf{Statement:}
During the preparation of this work the authors used GPT-5 (OpenAI) in order to improve
readability and clarity of the manuscript. After using this tool/service, the authors reviewed and edited the content as needed and take full responsibility for the content of the published article.

\printbibliography

@book{saleh_fundamentals_2018,
	address = {New Delhi},
	edition = {2nd. ed},
	title = {Fundamentals of photonics},
	isbn = {978-81-265-3774-7},
	language = {eng},
	publisher = {Wiley},
	author = {Saleh, Bahaa E.A.},
	collaborator = {Teich, Malvin Carl},
	year = {2018},
	note = {OCLC: 1378459105},
}

@article{roux_mid-infrared_2015,
	title = {Mid-infrared characterization of refractive indices and propagation losses in {GaSb}/{Al}$_{\textrm{x}}${Ga}$_{\textrm{1−x}}${AsSb} waveguides},
	volume = {107},
	issn = {0003-6951, 1077-3118},
	url = {https://pubs.aip.org/apl/article/107/17/171901/28700/Mid-infrared-characterization-of-refractive},
	doi = {10.1063/1.4934702},
	language = {en},
	number = {17},
	urldate = {2025-08-06},
	journal = {Applied Physics Letters},
	author = {Roux, S. and Barritault, P. and Lartigue, O. and Cerutti, L. and Tournié, E. and Gérard, B. and Grisard, A.},
	month = oct,
	year = {2015},
	pages = {171901},
}

@article{ferrini_optical_1998,
	title = {Optical functions from 0.02 to 6 {eV} of {Al}$_{\textrm{x}}${Ga}$_{\textrm{1−x}}${Sb}/{GaSb} epitaxial layers},
	volume = {84},
	issn = {0021-8979, 1089-7550},
	url = {https://pubs.aip.org/jap/article/84/8/4517/363549/Optical-functions-from-0-02-to-6-eV-of-AlxGa1-xSb},
	doi = {10.1063/1.368677},
	language = {en},
	number = {8},
	urldate = {2025-08-06},
	journal = {Journal of Applied Physics},
	author = {Ferrini, R. and Patrini, M. and Franchi, S.},
	month = oct,
	year = {1998},
	pages = {4517--4524},
}

@article{paskov_refractive_1997,
	title = {Refractive indices of {InSb}, {InAs}, {GaSb}, {InAs}$_{\textrm{x}}${Sb}$_{\textrm{1−x}}$, and {In}$_{\textrm{1−x}}${Ga}$_{\textrm{x}}${Sb}: {Effects} of free carriers},
	volume = {81},
	issn = {0021-8979, 1089-7550},
	shorttitle = {Refractive indices of {InSb}, {InAs}, {GaSb}, {InAsxSb1}−x, and {In1}−{xGaxSb}},
	url = {https://pubs.aip.org/jap/article/81/4/1890/515398/Refractive-indices-of-InSb-InAs-GaSb-InAsxSb1-x},
	doi = {10.1063/1.365360},
	language = {en},
	number = {4},
	urldate = {2025-08-07},
	journal = {Journal of Applied Physics},
	author = {Paskov, P. P.},
	month = feb,
	year = {1997},
	pages = {1890--1898},
}

@article{alibert_refractive_1991,
	title = {Refractive indices of {AlSb} and {GaSb}-lattice-matched {Al}$_{\textrm{x}}${Ga}$_{\textrm{1−x}}${As}$_{\textrm{y}}${Sb}$_{\textrm{1−y}}$ in the transparent wavelength region},
	volume = {69},
	issn = {0021-8979, 1089-7550},
	url = {https://pubs.aip.org/jap/article/69/5/3208/20681/Refractive-indices-of-AlSb-and-GaSb-lattice},
	doi = {10.1063/1.348538},
	language = {en},
	number = {5},
	urldate = {2025-08-06},
	journal = {Journal of Applied Physics},
	author = {Alibert, C. and Skouri, M. and Joullie, A. and Benouna, M. and Sadiq, S.},
	month = mar,
	year = {1991},
	pages = {3208--3211},
}

@article{adachi_optical_1989,
	title = {Optical dispersion relations for {GaP}, {GaAs}, {GaSb}, {InP}, {InAs}, {InSb}, {Al}$_{\textrm{x}}${Ga}$_{\textrm{1−x}}${As}, and {In}$_{\textrm{1−x}}${Ga}$_{\textrm{x}}${As}$_{\textrm{y}}${P}$_{\textrm{1−y}}$},
	volume = {66},
	issn = {0021-8979, 1089-7550},
	url = {https://pubs.aip.org/jap/article/66/12/6030/17370/Optical-dispersion-relations-for-GaP-GaAs-GaSb-InP},
	doi = {10.1063/1.343580},
	language = {en},
	number = {12},
	urldate = {2025-08-06},
	journal = {Journal of Applied Physics},
	author = {Adachi, Sadao},
	month = dec,
	year = {1989},
	pages = {6030--6040},
}

@article{schuchter_2-m_2024,
	title = {2-\textit{μ}m 1.5-{W} {Optically} {Pumped} {Semiconductor} {Membrane} {Laser}},
	volume = {36},
	copyright = {https://creativecommons.org/licenses/by/4.0/legalcode},
	issn = {1041-1135, 1941-0174},
	url = {https://ieeexplore.ieee.org/document/10472485/},
	doi = {10.1109/LPT.2024.3377261},
	language = {en},
	number = {8},
	urldate = {2025-09-15},
	journal = {IEEE Photonics Technology Letters},
	author = {Schuchter, Maximilian C. and Huwyler, Nicolas and Golling, Matthias and Gaulke, Marco and Keller, Ursula},
	month = apr,
	year = {2024},
	pages = {543--546},
}

@article{gaulke_optically_2025,
	title = {Optically {Pumped} {GaSb}-{Based} {Thin}-{Disk} {Laser} {Design} {Considerations} for {CW} and {Dual}-{Comb} {Operation} at a {Center} {Wavelength} {Around} 2 \textit{μ}m},
	volume = {31},
	copyright = {https://creativecommons.org/licenses/by/4.0/legalcode},
	issn = {1077-260X, 1558-4542},
	url = {https://ieeexplore.ieee.org/document/10664454/},
	doi = {10.1109/JSTQE.2024.3454521},
	number = {2: Pwr. and Effic. Scaling in},
	urldate = {2025-10-09},
	journal = {IEEE Journal of Selected Topics in Quantum Electronics},
	author = {Gaulke, Marco and Schuchter, Maximilian C. and Huwyler, Nicolas and Golling, Matthias and Willenberg, Benjamin and Phillips, Christopher R. and Keller, Ursula},
	month = mar,
	year = {2025},
	pages = {1--14},
}

@article{zia_high-power_2019,
	title = {High-power single mode {GaSb}-based 2 \textit{μ}m superluminescent diode with double-pass gain},
	volume = {115},
	issn = {0003-6951, 1077-3118},
	url = {https://pubs.aip.org/apl/article/115/23/231106/37479/High-power-single-mode-GaSb-based-2-m},
	doi = {10.1063/1.5127407},
	language = {en},
	number = {23},
	urldate = {2025-10-09},
	journal = {Applied Physics Letters},
	author = {Zia, Nouman and Viheriälä, Jukka and Koivusalo, Eero and Guina, Mircea},
	month = dec,
	year = {2019},
	pages = {231106},
}

@book{fujiwara_spectroscopic_2009,
	address = {Chichester},
	edition = {Printed with corrections},
	title = {Spectroscopic ellipsometry: principles and applications},
	isbn = {978-0-470-01608-4},
	shorttitle = {Spectroscopic ellipsometry},
	language = {eng},
	publisher = {Wiley},
	author = {Fujiwara, Hiroyuki},
	year = {2009},
}

@article{bett_gasb_2003,
	title = {{GaSb} photovoltaic cells for applications in {TPV} generators},
	volume = {18},
	issn = {0268-1242, 1361-6641},
	url = {https://iopscience.iop.org/article/10.1088/0268-1242/18/5/307},
	doi = {10.1088/0268-1242/18/5/307},
	language = {en},
	number = {5},
	urldate = {2025-11-18},
	journal = {Semiconductor Science and Technology},
	author = {Bett, A W and Sulima, O V},
	month = may,
	year = {2003},
	pages = {S184--S190},
}

@article{kwei-wei_huang_active_2013,
	title = {Active and passive infrared imager based on short-wave and mid-wave type-{II} superlattice dual-band detectors},
	volume = {38},
	copyright = {https://doi.org/10.1364/OA\_License\_v1\#VOR},
	issn = {0146-9592, 1539-4794},
	url = {https://opg.optica.org/abstract.cfm?URI=ol-38-1-22},
	doi = {10.1364/OL.38.000022},
	language = {en},
	number = {1},
	urldate = {2025-11-18},
	journal = {Optics Letters},
	author = {Kwei-wei Huang, Edward and Haddadi, Abbas and Chen, Guanxi and Hoang, Anh-Minh and Razeghi, Manijeh},
	month = jan,
	year = {2013},
	pages = {22},
}

@article{cerutti_gasb-based_2009,
	title = {{GaSb}-based {VCSELs} emitting in the mid-infrared wavelength range (2–3μm) grown by {MBE}},
	volume = {311},
	copyright = {https://www.elsevier.com/tdm/userlicense/1.0/},
	issn = {00220248},
	url = {https://linkinghub.elsevier.com/retrieve/pii/S0022024808012426},
	doi = {10.1016/j.jcrysgro.2008.11.026},
	language = {en},
	number = {7},
	urldate = {2025-10-09},
	journal = {Journal of Crystal Growth},
	author = {Cerutti, L. and Ducanchez, A. and Narcy, G. and Grech, P. and Boissier, G. and Garnache, A. and Tournié, E. and Genty, F.},
	month = mar,
	year = {2009},
	pages = {1912--1916},
}

@article{joullie_gasb-based_2003,
	title = {{GaSb}-based mid-infrared 2–5 μm laser diodes},
	volume = {4},
	issn = {1878-1535},
	url = {https://comptes-rendus.academie-sciences.fr/physique/articles/10.1016/S1631-0705(03)00098-7/},
	doi = {10.1016/S1631-0705(03)00098-7},
	language = {en},
	number = {6},
	urldate = {2025-10-09},
	journal = {Comptes Rendus. Physique},
	author = {Joullié, André and Christol, Philippe},
	month = jul,
	year = {2003},
	pages = {621--637},
}

@article{yasuda_growth_2024,
	title = {Growth of {InGaSb}/{AlInGaSb} quantum cascade laser structures on {GaSb} substrates by molecular beam epitaxy},
	volume = {636},
	issn = {00220248},
	url = {https://linkinghub.elsevier.com/retrieve/pii/S0022024824001556},
	doi = {10.1016/j.jcrysgro.2024.127720},
	language = {en},
	urldate = {2025-10-09},
	journal = {Journal of Crystal Growth},
	author = {Yasuda, Hiroaki and Sekine, Norihiko and Hosako, Iwao},
	month = jun,
	year = {2024},
	pages = {127720},
}

@article{schuchter_composition-controlled_2025,
	title = {Composition-controlled recovery time of {SWIR} (2–2.4 µm) {GaSb}-based {SESAMs}},
	volume = {33},
	issn = {1094-4087},
	url = {https://opg.optica.org/abstract.cfm?URI=oe-33-7-14750},
	doi = {10.1364/OE.558108},
	language = {en},
	number = {7},
	urldate = {2025-10-09},
	journal = {Optics Express},
	author = {Schuchter, Maximilian C. and Gaulke, Marco and Huwyler, Nicolas and Golling, Matthias and Guina, Mircea and Keller, Ursula},
	month = apr,
	year = {2025},
	pages = {14750},
}

@article{alaydin_bandgap_2022,
	title = {Bandgap engineering, monolithic growth, and operation parameters of {GaSb}-based {SESAMs} in the 2–2.4 µm range},
	volume = {12},
	issn = {2159-3930},
	url = {https://opg.optica.org/abstract.cfm?URI=ome-12-6-2382},
	doi = {10.1364/OME.459232},
	language = {en},
	number = {6},
	urldate = {2025-10-09},
	journal = {Optical Materials Express},
	author = {Alaydin, B. Özgür and Gaulke, Marco and Heidrich, Jonas and Golling, Matthias and Barh, Ajanta and Keller, Ursula},
	month = jun,
	year = {2022},
	pages = {2382},
}

@article{nelder_simplex_1965,
	title = {A {Simplex} {Method} for {Function} {Minimization}},
	volume = {7},
	issn = {0010-4620, 1460-2067},
	url = {https://academic.oup.com/comjnl/article-lookup/doi/10.1093/comjnl/7.4.308},
	doi = {10.1093/comjnl/7.4.308},
	language = {en},
	number = {4},
	urldate = {2025-10-02},
	journal = {The Computer Journal},
	author = {Nelder, J. A. and Mead, R.},
	month = jan,
	year = {1965},
	pages = {308--313},
}

@article{gaulke_high_2021,
	title = {High average output power from a backside-cooled 2-µm {InGaSb} {VECSEL} with full gain characterization},
	volume = {29},
	issn = {1094-4087},
	url = {https://opg.optica.org/abstract.cfm?URI=oe-29-24-40360},
	doi = {10.1364/OE.438157},
	language = {en},
	number = {24},
	urldate = {2025-09-15},
	journal = {Optics Express},
	author = {Gaulke, Marco and Heidrich, Jonas and Özgür Alaydin, B. and Golling, Matthias and Barh, Ajanta and Keller, Ursula},
	month = nov,
	year = {2021},
	pages = {40360},
}

@article{huwyler_3-w_2023,
	title = {3-{W} output power from a 2-µm {InGaSb} {VECSEL} using a hybrid metal-semiconductor {Bragg} reflector},
	volume = {13},
	issn = {2159-3930},
	url = {https://opg.optica.org/abstract.cfm?URI=ome-13-3-833},
	doi = {10.1364/OME.485694},
	language = {en},
	number = {3},
	urldate = {2025-09-15},
	journal = {Optical Materials Express},
	author = {Huwyler, Nicolas and Gaulke, Marco and Heidrich, Jonas and Golling, Matthias and Barh, Ajanta and Keller, Ursula},
	month = mar,
	year = {2023},
	pages = {833},
}

@article{luce_tmm-fast_2022,
	title = {{TMM}-{Fast}, a transfer matrix computation package for multilayer thin-film optimization: tutorial},
	volume = {39},
	issn = {1084-7529, 1520-8532},
	shorttitle = {{TMM}-{Fast}, a transfer matrix computation package for multilayer thin-film optimization},
	url = {https://opg.optica.org/abstract.cfm?URI=josaa-39-6-1007},
	doi = {10.1364/JOSAA.450928},
	language = {en},
	number = {6},
	urldate = {2025-08-25},
	journal = {Journal of the Optical Society of America A},
	author = {Luce, Alexander and Mahdavi, Ali and Marquardt, Florian and Wankerl, Heribert},
	month = jun,
	year = {2022},
	pages = {1007},
}

@misc{lebigot_uncertainties_2024,
	title = {Uncertainties: a {Python} package for calculations with uncertainties},
	url = {http://pythonhosted.org/uncertainties/},
	author = {Lebigot, Eric O.},
	month = jul,
	year = {2024},
}

@article{skauli_improved_2003,
	title = {Improved dispersion relations for {GaAs} and applications to nonlinear optics},
	volume = {94},
	issn = {0021-8979, 1089-7550},
	url = {https://pubs.aip.org/jap/article/94/10/6447/761848/Improved-dispersion-relations-for-GaAs-and},
	doi = {10.1063/1.1621740},
	language = {en},
	number = {10},
	urldate = {2025-08-21},
	journal = {Journal of Applied Physics},
	author = {Skauli, T. and Kuo, P. S. and Vodopyanov, K. L. and Pinguet, T. J. and Levi, O. and Eyres, L. A. and Harris, J. S. and Fejer, M. M. and Gerard, B. and Becouarn, L. and Lallier, E.},
	month = nov,
	year = {2003},
	pages = {6447--6455},
}

@article{perner_simultaneous_2023,
	title = {Simultaneous measurement of mid-infrared refractive indices in thin-film heterostructures: {Methodology} and results for {GaAs}/{AlGaAs}},
	volume = {5},
	issn = {2643-1564},
	shorttitle = {Simultaneous measurement of mid-infrared refractive indices in thin-film heterostructures},
	url = {https://link.aps.org/doi/10.1103/PhysRevResearch.5.033048},
	doi = {10.1103/PhysRevResearch.5.033048},
	language = {en},
	number = {3},
	urldate = {2024-02-15},
	journal = {Physical Review Research},
	author = {Perner, Lukas W. and Truong, Gar-Wing and Follman, David and Prinz, Maximilian and Winkler, Georg and Puchegger, Stephan and Cole, Garrett D. and Heckl, Oliver H.},
	month = jul,
	year = {2023},
	pages = {033048},
}

@misc{newville_lmfit_2025,
	title = {{LMFIT}: {Non}-{Linear} {Least}-{Squares} {Minimization} and {Curve}-{Fitting} for {Python}},
	copyright = {Creative Commons Attribution 4.0 International},
	shorttitle = {{LMFIT}},
	url = {https://zenodo.org/doi/10.5281/zenodo.15014437},
	doi = {10.5281/ZENODO.15014437},
	urldate = {2025-08-06},
	publisher = {Zenodo},
	author = {Newville, Matthew and Otten, Renee and Nelson, Andrew and Stensitzki, Till and Ingargiola, Antonino and Allan, Daniel and Fox, Austin and Carter, Faustin and Rawlik, Michal},
	month = mar,
	year = {2025},
}

@article{wasiak_absorption_2018,
	title = {Absorption and dispersion in undoped epitaxial {GaSb} layer},
	volume = {5},
	issn = {2053-1591},
	url = {https://iopscience.iop.org/article/10.1088/2053-1591/aaae86},
	doi = {10.1088/2053-1591/aaae86},
	number = {2},
	urldate = {2025-08-06},
	journal = {Materials Research Express},
	author = {Wasiak, Michał and Motyka, Marcin and Smołka, Tristan and Ratajczak, Jacek and Jasik, Agata},
	month = feb,
	year = {2018},
	pages = {025907},
}

@article{terven_comprehensive_2025,
	title = {A comprehensive survey of loss functions and metrics in deep learning},
	volume = {58},
	issn = {1573-7462},
	url = {https://link.springer.com/10.1007/s10462-025-11198-7},
	doi = {10.1007/s10462-025-11198-7},
	language = {en},
	number = {7},
	urldate = {2025-08-06},
	journal = {Artificial Intelligence Review},
	author = {Terven, Juan and Cordova-Esparza, Diana-Margarita and Romero-González, Julio-Alejandro and Ramírez-Pedraza, Alfonso and Chávez-Urbiola, E. A.},
	month = apr,
	year = {2025},
	pages = {195},
}

@article{munoz_uribe_near-band-gap_1996,
	title = {Near-band-gap refractive index of {GaSb}},
	volume = {38},
	copyright = {https://www.elsevier.com/tdm/userlicense/1.0/},
	issn = {09215107},
	url = {https://linkinghub.elsevier.com/retrieve/pii/0921510795015019},
	doi = {10.1016/0921-5107(95)01501-9},
	language = {en},
	number = {3},
	urldate = {2025-08-06},
	journal = {Materials Science and Engineering: B},
	author = {Muñoz Uribe, M. and Miranda, R.S. and Zakia, M.B. and De Souza, C.F. and Ribeiro, C.A. and Clerice, J.H. and Patel, N.B.},
	month = apr,
	year = {1996},
	pages = {259--262},
}

@article{linnik_calculations_2002,
	title = {Calculations of optical properties for quaternary {III}–{V} semiconductor alloys in the transparent region and above (0.2–4.{0eV})},
	volume = {318},
	copyright = {https://www.elsevier.com/tdm/userlicense/1.0/},
	issn = {09214526},
	url = {https://linkinghub.elsevier.com/retrieve/pii/S0921452602004672},
	doi = {10.1016/S0921-4526(02)00467-2},
	language = {en},
	number = {2-3},
	urldate = {2025-08-06},
	journal = {Physica B: Condensed Matter},
	author = {Linnik, M. and Christou, A.},
	month = jun,
	year = {2002},
	pages = {140--161},
}

@article{lasdon_adaptive_2010,
	title = {Adaptive memory programming for constrained global optimization},
	volume = {37},
	copyright = {https://www.elsevier.com/tdm/userlicense/1.0/},
	issn = {03050548},
	url = {https://linkinghub.elsevier.com/retrieve/pii/S0305054809002937},
	doi = {10.1016/j.cor.2009.11.006},
	language = {en},
	number = {8},
	urldate = {2025-08-06},
	journal = {Computers \& Operations Research},
	author = {Lasdon, Leon and Duarte, Abraham and Glover, Fred and Laguna, Manuel and Martí, Rafael},
	month = aug,
	year = {2010},
	pages = {1500--1509},
}

@article{kim_modeling_1992,
	title = {Modeling the optical dielectric function of semiconductors: {Extension} of the critical-point parabolic-band approximation},
	volume = {45},
	copyright = {http://link.aps.org/licenses/aps-default-license},
	issn = {0163-1829, 1095-3795},
	shorttitle = {Modeling the optical dielectric function of semiconductors},
	url = {https://link.aps.org/doi/10.1103/PhysRevB.45.11749},
	doi = {10.1103/PhysRevB.45.11749},
	language = {en},
	number = {20},
	urldate = {2025-08-06},
	journal = {Physical Review B},
	author = {Kim, Charles C. and Garland, J. W. and Abad, H. and Raccah, P. M.},
	month = may,
	year = {1992},
	pages = {11749--11767},
}

@article{gao_implementing_2012,
	title = {Implementing the {Nelder}-{Mead} simplex algorithm with adaptive parameters},
	volume = {51},
	copyright = {http://www.springer.com/tdm},
	issn = {0926-6003, 1573-2894},
	url = {http://link.springer.com/10.1007/s10589-010-9329-3},
	doi = {10.1007/s10589-010-9329-3},
	language = {en},
	number = {1},
	urldate = {2025-08-06},
	journal = {Computational Optimization and Applications},
	author = {Gao, Fuchang and Han, Lixing},
	month = jan,
	year = {2012},
	pages = {259--277},
}

@article{aspnes_dielectric_1983,
	title = {Dielectric functions and optical parameters of {Si}, {Ge}, {GaP}, {GaAs}, {GaSb}, {InP}, {InAs}, and {InSb} from 1.5 to 6.0 {eV}},
	volume = {27},
	copyright = {http://link.aps.org/licenses/aps-default-license},
	issn = {0163-1829},
	url = {https://link.aps.org/doi/10.1103/PhysRevB.27.985},
	doi = {10.1103/PhysRevB.27.985},
	language = {en},
	number = {2},
	urldate = {2025-08-06},
	journal = {Physical Review B},
	author = {Aspnes, D. E. and Studna, A. A.},
	month = jan,
	year = {1983},
	pages = {985--1009},
}

@article{djurisic_modeling_2000,
	title = {Modeling the optical properties of {AlSb}, {GaSb}, and {InSb}},
	volume = {70},
	copyright = {http://www.springer.com/tdm},
	issn = {0947-8396, 1432-0630},
	url = {http://link.springer.com/10.1007/s003390050006},
	doi = {10.1007/s003390050006},
	number = {1},
	urldate = {2025-08-06},
	journal = {Applied Physics A: Materials Science \& Processing},
	author = {Djurišić, A.B. and Li, E.H. and Rakić, D. and Majewski, M.L.},
	month = jan,
	year = {2000},
	pages = {29--32},
}
\end{document}